\documentclass[preprint,12pt]{elsarticle}   

\usepackage[]{graphicx}
\usepackage{amssymb,amsmath}
\usepackage{bm}
\usepackage{color}

\journal{Colloids and Surfaces B}

\begin{document} 
\begin{frontmatter}

\title{Surface charge and hydrodynamic coefficient measurements of {\it Bacillus subtilis} spore by Optical Tweezers}

\author[a]{Giuseppe Pesce}
\ead{giuseppe.pesce@fisica.unina.it, Tel. ++30081676274}
\author[a]{Giulia Rusciano}
\author[a,c]{Antonio Sasso}
\author[b]{Rachele Isticato}
\author[b]{Teja Sirec}
\author[b]{Ezio Ricca}

\address[a]{Dipartimento di Fisica Universit\`a degli studi di Napoli, Complesso Universitario Monte S. Angelo, Via Cintia 80126, Napoli, Italy}
\address[c]{CNR Istituto Nazionale di Ottica - Sezione di Napoli, Via Campi Flegrei, 34 80078 Pozzuoli (Napoli), Italy}
\address[b]{Dipartimento di Biologia Universit\`a degli studi di Napoli, Complesso Universitario Monte S. Angelo, Via Cintia 80126, Napoli, Italy}

\begin{abstract} 
In this work we report on the simultaneous measurement of the hydrodynamic coefficient and the electric 
charge of single {\it Bacillus subtilis} spores. The latter has great importance in protein binding to spores and 
in the adhesion of spores onto surfaces. The charge and the hydrodynamic coefficient were measured by 
an accurate procedure based on the analysis of the motion of single spores confined by an optical trap. 
The technique has been validated using charged spherical polystyrene beads. The excellent agreement 
of our results with the expected values demonstrates the quality of our procedure. We measured the 
charge of spores of {\it B. subtilis} purified from a wild type strain and from two isogenic mutants characterized by 
an altered spore surface. Our technique is able to discriminate the three spore types used, by their charge 
and by their hydrodynamic coefficient which is related to the hydrophobic properties of the spore surface.

\end{abstract}

\end{frontmatter} 
\section{INTRODUCTION}
\label{sec:intro}  
The electrostatic interaction among charged colloids or among colloids and surfaces is one of fundamental issues in condensed matter physics. For instance, the electrostatic interactions between charged microparticles determine the stability of colloidal dispersions and particle aggregation. In biology, these electrostatic forces are crucial for the interpretation of ion adsorption and ion permeation processes as well as in adhesion processes of bacteria and bacterial spores onto surfaces. These processes are the base of a variety of phenomena ranging from the development of new biomaterials for biomedical application to the prevention of food contamination by bacterial biofilms and to the delivery of drugs and antigens to human mucosal surfaces \cite{CuttingIRI09}.
In this context the adhesion of bacterial spores to surfaces is of particular interest for at least two reasons: (i) decontamination of spores during the industrial food chain (from food preparation to packaging) and (ii) use of spores as a platform to deliver drugs and antigens to the human mucosal surfaces.

Spores are metabolically quiescent cells, extremely resistant to harsh environmental conditions. Spore resistance to extreme con- ditions is due to their structure characterized by a dehydrated cytoplasm surrounded by several protective layers: the thick cortex formed of peptidoglycan, a multilayered coat mostly formed by proteins and the recently identified crust, formed of proteins and glycoproteins \cite{McKenneyNRM13}. Because of their resistance to thermal and 
chemical treatments treatments spores cannot be efficiently removed by standard procedures. In the presence of water and organic molecules spores germinate originating cells able to grow and compromise the quality and safety of food products. Understanding the physicochemical nature of the interaction between spores and surface could probably lead to the use of more appropriate decontamination procedures. The resistance properties of spores, together with the safety record of several spore-formers species, is also the base for considering them as a mucosal delivery system  \cite{CuttingIRI09}. Also in this case, a detailed knowledge of the 
surface properties of  spores would help to rationalize the display of drugs and antigens on the spore surface 
 \cite{HuangV10,SirecMCF12,IsticatoMCF13}.

The interaction potential between colloids and between colloids and surfaces is intensively discussed in 
literature and remains a challenge for both experiment and theory. Despite its limitations, the Derjaguin-Landau-
Verwey-Overbeek (DLVO) theory \cite{Israelachvili11} describes very well the interaction between charged 
colloids in aqueous salt solutions.

Therefore the charge carried by a single colloid or biological object is a very relevant parameter to be measured. 
Usually this can be done by measuring the electrophoretic mobility $\mu$, that is defined as the ratio of the 
velocity $v$ of a charged particle over the applied electric field strength $E$ i.e. $\mu=v/E$. The zeta potential ($
\zeta$) is the electrokinetic potential measured at the slipping 
plane in the double layer and is related to the electrophoretic mobility by the relation:
$\mu=\epsilon_r\epsilon_0 \zeta/\eta$, being $\eta$ the viscosity of the fluid and $\epsilon_r$ and $\epsilon_0$ 
the dielectric constant and the absolute permittivity of vacuum, respectively. Therefore the knowledge of these 
associated parameters is of fundamental importance to characterise a colloidal system.
Commercial instruments, {\em Zetasizers}, are today available to measure the electrophoretic mobility of 
microscopic particles and, consequently, the zeta potential. These instruments are based on Laser Doppler 
Velocimetry (LDV) and Phase Analysis Light Scattering (PALS) techniques. Nevertheless these instruments give 
ensemble averaged results which have been reported to be dependent on the particle concentration \cite
{ReiberJCIS07} and are not adequate when the shape of particles differs significantly from that of a sphere as in 
the case of the bacterial spores investigated in this work.

In the last three decades, optical tweezers \cite{AshkinOL86,AshkinPNAS97} (OT), have revealed as a formidable tool in many areas of science. In particular, they have been used to shed light in colloidal physics for their ability to manipulate 
single particles. They are based on a strongly focussed laser beam that exerts a restoring force on a microscopic dielectric particle 
close to the focal point. In such a way the trapped particle results to be confined in an optical potential that can be considered with 
a good approximation harmonic. So far, the restoring force is elastic: $F_{opt}=-\kappa x$, being $\kappa$ the stiffness of the 
optical trap and $x$ the displacement from the laser focus.  Optical Tweezers in combination with a nanometre resolution position 
detection technique \cite{GittesOL98} represent a very sensitive force transducer allowing the measurement of forces from 
hundredths of pN down to few fN  \cite{RohrbachPRL05, ImparatoPRE07} . In particular, recent experiments have demonstrated the possibility to measure 
electrophoretic mobility of single colloids in polar and non polar fluids \cite{StrubbePRL07,SethJCP07,SemenovJCIS09} .

Here we show how to extend these techniques to the case of single bacterial spore dispersed in water.  
In particular we demonstrate that our technique is able to measure, simultaneously, the hydrodynamic coefficient and the electric charge of single particles with arbitrary shape. In this experiment we used bacterial spores that, at first approximation, can be considered of ellipsoidal shape and with a complex surface. The only requirement of our technique is that the confined thermal motion of the trapped object is purely translational. When this condition is fulfilled the hydrodynamic coefficient and the electric charge can be measured in a fast, reliable and precise way.

In addition we demonstrate the possibility to discriminate genetically modified bacillus spores from their surface charge.
Our method is also able to estimate the hydrodynamic factor of wild type and mutants spores. This could open the possibility to measure the degree of hydrophobicity of spores in different environmental conditions.

\section{Theory}
\label{sec.theory}
The motion of a  charged particle of diameter $d$ and mass $m$ subjected to a restoring force with stiffness $\kappa$  
and to an external electric field is described by the Langevin equation (here for one coordinate only):
\begin{equation}
m\ddot{x}(t)+\gamma \dot{x}(t) +\kappa x= F_{T}(t)+ F_E
\label{eq.lang}
\end{equation}
where $\gamma$ s the hydrodynamic coefficient which, for a spherical particle, is related to the medium viscosity by the
relation: 
$\gamma=3\pi\eta d$. $F_{T}(t)$ is the random thermal force that drives the Brownian motion, it has the 
statistical properties of white noise and is defined as:
\begin{equation}
F_{T}(t)=\sqrt{2\gamma k_B T}\xi(t)
\end{equation}
where $k_B$ is the Boltzmann constant, $T$ the absolute temperature and $\xi(t)$ is the zero-mean, 
$\delta$-correlated Gaussian white noise.

In a fluid with low Reynolds number, like water, the inertial term can be neglected since the decay time due to viscous 
losses is of the order of $m/\gamma$, i.e. $10^{-7}\,s$ which is much smaller than the time resolution used in this 
experiment (of the order of $10^{-5}\,s$, see below). 

The electric force is $F_E=Q_{eff}\cdot E$ where $Q_{eff}$ is the effective charge of the particle, resulting from the 
screening of free ions, and $E$ the  electric field strength. 
In this experiment we used a sinusoidal electric field $E(t)=E_0sin(2\pi f_p t)$, being $f_p$ the modulation 
frequency, thus Eq. (\ref{eq.lang}) becomes:
\begin{equation}
\gamma \dot{x}(t) +\kappa x= F_{T}(t)+ Q_{eff}E_0sin(2\pi f_p t)
\label{eq.langE}
\end{equation}

Since this equation is linear, its solution is simply the sum of two terms $x(t)=x_T(t)+x_E(t)$: the first represents the 
motion of the particle in presence of the thermal motion only, while the second  is due to the electric field only \cite
{SethJCP07}. In  particular the latter is the well known deterministic motion of a driven oscillator:
\begin{equation}
x_E(t)=\frac{Q_{eff}E_0}{\kappa\left[ 1+( f_p/f_C)^2\right]^{1/2}}sin(2\pi f_p t-\phi)
\end{equation}
where $f_C=\kappa/(2\pi\gamma)$ is the corner frequency of the optical trap and $\phi$ is a constant phase lag 
between the particle motion and the driving field: $\phi=arctg( f_p/f_C)$.

It is very convenient to study this dynamics using the AutoCorrelation Function (ACF) defined as $ac(t)=<x(t')x(t' + t)>$.  
Since the periodic electric field and the thermal motion are uncorrelated, the total autocorrelation function is, again, 
the sum of two autocorrelation functions: one for the thermal and another for the periodic motion induced by the electric field:
\begin{equation}
ac(t)=ac_{T}(t)+ac_{E}(t) = \\ 
\frac{k_B T}{\kappa}e^{-t/\tau}+ \frac{Q_{eff}E_0}{2\kappa^2\left[ 1+( f_p/f_C)^2\right]^
{1/2}}cos(2\pi f_p t)
\label{eq.ac}
\end{equation}
where $\tau=\gamma/\kappa$ is the characteristic time of the optical trap which is related to the 
{\em corner frequency} $f_C=1/(2\pi \tau)$. Equation (\ref{eq.ac}) can be written as the normalized function $acn(t)=ac(t)/ac(0)$:
\begin{equation}
acn(t)=\frac{1}{1+\Gamma^2}e^{-t/\tau}+ \frac{\Gamma^2}{1+\Gamma^2}cos(2\pi f_p t)
\label{eq.acn}
\end{equation}

The particle motion can be also described in the frequency domain through its Power Spectral Density (PSD). The Wiener-Khinchin 
theorem states that the spectral decomposition of the autocorrelation function is given by the power spectrum of that process, 
therefore the PSD is given by:
\begin{equation}
S(f)=\frac{A_{psd}}{f_C^2+f^2}+B_{psd}\cdot \delta(f- f_p) = \frac{k_BT}{\pi^2\gamma}\frac{1}{f_C^2+f^2}+
\frac{k_BT}{\kappa}\pi\Gamma^2\delta(f- f_p)
\label{eq.psd}
\end{equation}
which is the superposition of a lorentzian function related to the thermal motion and a peak at the driving frequency of the forcing electric field.

The parameter $\Gamma$ that appears in Eqs.\ref{eq.acn} and \ref{eq.psd} is the ratio of the periodic force and the thermal force scaled by the ratio of the frequency 
modulation of the external field and the corner frequency of the optical trap:
\begin{equation}
\Gamma^2=\frac{<F_E^2>/<F_{T}^2>}{1+( f_p/f_C)^2}
\label{eq.gammae}
\end{equation}
For weak electric fields $\Gamma \ll $ and the particle motion is dominated by the thermal fluctuation, while for strong 
fields $\Gamma \gg 1$ and the motion is essentially periodic and the thermal fluctuations can be regarded as small 
perturbation.
$\Gamma$ is readily obtained from the analysis of the normalized autocorrelation function and it can directly give the 
value of the charge carried by the trapped particle. In fact the mean square thermal force is $<F_T^2>=k_BT/\kappa$, 
while that of the periodic field is  $<F_E^2>=Q^2_{eff}E^2/2$. Therefore, once $\Gamma$ is determined from Eq.\ref
{eq.acn} the effective charge  $Q_{eff}$ can be deduced from the following equation:
\begin{equation}
Q_{eff}=\frac{\Gamma}{E}\sqrt{4\pi\gamma k_BT\frac{(f_C^2+ f_p^2)}{f_C}}
\label{eq.charge}
\end{equation}

The effective charge of the trapped particle is determined by experimental parameters ($\Gamma$, $f_C$) obtained 
from the analysis of its motion in the optical potential well and also on the knowledge of the hydrodynamic coefficient $
\gamma$. We remember that $\gamma$ depends on the geometric form factor of the particle and on the viscosity of the fluid. 
The form factor is well known for spherical particles and it is equal to $3\pi\eta d$. 
For ellipsoidal particles that move along one of their principal axes the geometric form factors are 
equal to those of a sphere multiplied by a correction coefficient $\Phi$  \cite{SCIHan06,PREHan09}. In the specific case of an 
ellipsoid with equal short axes the coefficients are:
\begin{equation}
\begin{split}
\Phi_a=\frac{8}{3}\frac{1}{\left[ \frac{2r}{1-r^2}+\frac{2r^2-1}{(r^2-1)^{3/2}}ln\left( \frac{r+\sqrt{r^2-1}}{r-\sqrt
{r^2-1}}\right)   \right]}, \\
\Phi_b=\frac{8}{3}\frac{1}{\left[ \frac{r}{r^2-1}+\frac{2r^2-3}{(r^2-1)^{3/2}}ln\left( r+\sqrt{r^2-1} \right)   \right]}
\end{split}
\label{eq.gfactor}
\end{equation}
being $r$ the aspect ratio of the ellipsoid ($r=a/b$ with $a$ and $b$ the long and short diameters). 
Unfortunately for biological particles, like bacterial spores, the hydrodynamic coefficient is unknown and, as we will show below, 
the approximation to particles with similar shape (mostly ellipsoidal) fails. Therefore to overcome this issue, it is necessary 
to measure the hydrodynamic factor directly.

In OT the position $x$ of the trapped particle is measured with a quadrant photodiode using the forward light 
scattering detection technique \cite{GittesOL98,BuoscioloOC04,SorensenRSI04}.
The output signal $V_x$ of the QPD should be converted in meter units (conversion factor $\beta$). Usually from the 
analysis the thermal contribution of the auto correlation function or, equivalently, of the PSD (see Eq. (\ref{eq.acn}) and \ref
{eq.psd}) the conversion factor $\beta=x/V_x$ is easily obtained provided $\gamma$ is a known quantity, i.e. the viscosity of the 
medium and the form factor are known  \cite{GittesOL98,BuoscioloOC04,SorensenRSI04} . 

However it is possible to measure the conversion factor $\beta$ and the hydrodynamic coefficient of a trapped particle 
simultaneously. The method is based on the measurement of the dynamics of a trapped particle when the fluid 
surrounding the particle is moved in a periodic way \cite{TolicRSI06}. 
This problem is formally identical to the case, discussed above, of a particle driven by an external periodic electric field. 
The motion of the particle is, again, the linear superposition of the Brownian thermal diffusion with the deterministic 
motion due to the dragging of the oscillating fluid.

Again, this oscillation results in a cosinusoidal component in the ACF or, equivalently, in a sharp peak in the PSD.
 The signal power under this peak can be evaluated theoretically 
and results equal to:
\begin{equation}
W_{th}=\frac{A^2 f_d^2}{2( f_d^2+f_C^2)}
\label{eq.peakpower}
\end{equation}
where $A$ is the amplitude of the fluid periodic motion, measured in meter and $ f_d$ the modulation frequency of 
the fluid. Since $W_{th}$  can be measured ($W_{exp}$)  the conversion factor $\beta$ is, therefore, determined by the square 
root of their ratio:
\begin{equation}
\beta=\sqrt{\frac{W_{th}}{W_{exp}}}= \frac{A  f_d}{\sqrt{2 W_{exp}[ f_d^2+ f_C^2]}}
\label{eq.beta}
\end{equation}

Once $\beta$ has been determined, the hydrodynamic coefficient is straightforwardly calculated 
from the amplitude of the experimental power spectral density $A_{psd}$, by the following equation:
\begin{equation}
\gamma=\frac{k_BT}{\pi^2 \beta^2  A_{psd}}.
\label{eq.gamma}
\end{equation}
$A_{psd}$ and the corner frequency $f_C$ are easily obtained from a fitting of the PSD itself. 
Therefore it is possible to measure the hydrodynamic coefficient directly without any information on the size and 
geometrical form of the trapped particle. The only assumption required is that the motion of particle should be only 
translational, i.e. the particle should not rotate. This condition is fulfilled for the case of the bacterial spores
used in this experiment as shown below.

\section{Materials and Methods}

\begin{figure}[t]
\begin{center}
\includegraphics[width=0.7\linewidth]{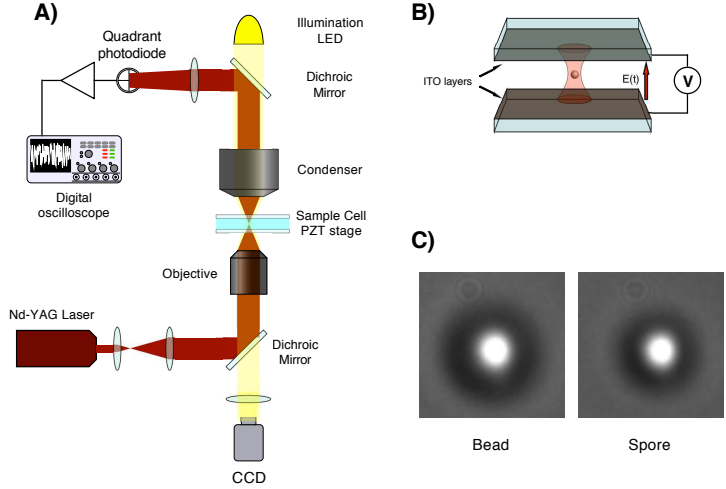}
\caption{(A) Schematic view of the optical tweezers setup, (B) a detail of the sample cell with ITO planar electrodes, (C) bright light image of a trapped bead (left) and {\it wild type} spore (right).}
\label{fig.setup}
\end{center}
\end{figure}

The experimental setup (schematically shown in Fig.\ref{fig.setup}) consists of an optical tweezers
build on a high-stability home-made optical microscope equipped with a high-numerical-aperture
water-immersion objective lens (Olympus, UPLAPO60XW3, NA=1.2). The optical trap is generated by a 
frequency and  amplitude stabilized Nd-YAG laser ($\mathrm{\lambda=1.064\,\mu m}$, $
\mathrm{500\,mW}$ maximum output power, Innolight Mephisto) \cite{PesceRSI05} . The laser beam was 
expanded to a final diameter of 10 mm to have a proper overfilling of the back aperture of the 
objective lens. The laser power was precisely regulated rotating a half-wave-plate before a polarizer.
To minimize external noise, the experimental setup was mounted on a passive vibration isolation optical table.  Moreover, 
the laser paths were kept as short as possible to avoid pointing fluctuations and the laser beam was enclosed in plastic 
pipes wherever possible.  Finally, the whole setup was enclosed in a thermal and acoustic insulated box to prevent air 
circulation and temperature drifts and to avoid heating, a white cold LED was used to illuminate the sample. The temperature 
inside the box was continuously monitored, during each set of measurements the temperature fluctuations were within 0.2 
$^{\circ} C$.

The 3D position of the trapped particle was monitored through the forward scattered light \cite{GittesOL98} imaged onto 
an InGaAs Quadrant Photodiode (QPD, Hamamatsu G6849) at the back focal plane of the condenser lens. 

For movements in the plane (x,y) perpendicular to the beam propagation axis (z)  the QPD-response was linear for 
displacements up to $\mathrm{300\, nm}$  with $\mathrm{2\, nm}$ resolution. The
z position is measured following the variations of the total laser intensity measured at the back focal plane of the 
condenser. For small displacements the laser intensity is a monotonic function of the axial particle position. 
Since the laser intensity was very stable ($<$ 0.1 \% rms) it was not necessary to normalize the three output signals by 
the intensity at the back focal plane. 
Output signals from the quadrant photodiode were acquired for 20 s at 50 kHz using a digital oscilloscope (Tektronix 
TDS5034B), then data were sent to a computer for further analysis.

We used negatively charged sulfate coated polystyrene micro-spheres (Postnova, $\mathrm{1.06\,g/cm^3}$ density, 
1.65 refractive index) with a diameter of $\mathrm{1.00\pm0.08}$ $\mathrm{\, \mu m}$. Particles were diluted in distilled deionized 
water to a final concentration of a few particles/$\mathrm{\mu l}$. The surface charge density, provided by the 
manufacturer, was 5.7 $\mathrm{\mu}C/cm^2$ that corresponds to a total charge of -1.79$\times 10^{-13}$ C.

The spores used in this experiment were purified from a standard laboratory strain of {\it B. subtilis} (PY79) and from isogenic mutants 
carrying null mutations in the {\it cotE (cotE::cm)}  \cite{ZhengGD88} or in the {\it cotH (cotH::spc)}   \cite{NaclerioJB96} gene. 

After 30 h of growth in Difco Sporulation medium (DSM) at $37\,^{\circ}C$ with vigorous shaking, spores were harvested, 
washed three times with distilled water and purified by gastrografin gradient as described in ref.s  \cite{Nicholson90,IsticatoJB08,SirecMCF12} . 
Spore counts were determined by serial dilution and plate-counting. Spores were diluted to a final concentration of about 
50$\div$100 spores/$\mu l$ immediately before use to prevent adhesion to vial surface and loaded into the sample cell. 

Before starting every measurement,  sulfate polystyrene beads or spores were dispersed in fresh milli-Q water 
at pH=6.5. After dilution we found about 1-5 spores/beads in the field of view of our microscope. 
A droplet  of such solution ($\mathrm{50\, \mu l}$) was 
placed inside a sample chamber made of a $150 \, \mu m$-thick coverslip and a microscope slide, both 
coated with ITO (Indium-Tin-Oxide) that act as planar electrodes. The sample cell was mounted on a 
closed-loop piezoelectric stage (Physik Instrumente PI-517.3CL), which allows movements with nanometer 
resolution. 

Since the experiment was performed in water which is a polar fluid it was necessary to evaluate the effects of the free 
ions. The first and the most relevant effect is the polarization of the electrodes due to the 
screening of the counterions. When a voltage difference is applied to the electrodes, the free ions present in the fluid are 
attracted toward them. The result is that ions screen the charge accumulated on the electrodes by the external voltage 
supply and the electric field between them vanishes.  A simple way to avoid this effect is to use alternating electric field. 
However due to the finite drift velocity of free ions, there is a dependence on the frequency of the alternating field: increasing the 
frequency the screening effect reduces since the free ions cannot reach the electrodes within half the time period, i.e. 
before inverting their motion, so the response of the charged particle increases accordingly. Nevertheless after a certain 
value of the frequency the response starts to decrease because the particle cannot follow the oscillating driving field 
due to the viscosity of medium (the viscous medium acts like a low pass filter for the particle motion).

\begin{figure}[h]
\begin{center}
\includegraphics[width=0.9\linewidth]{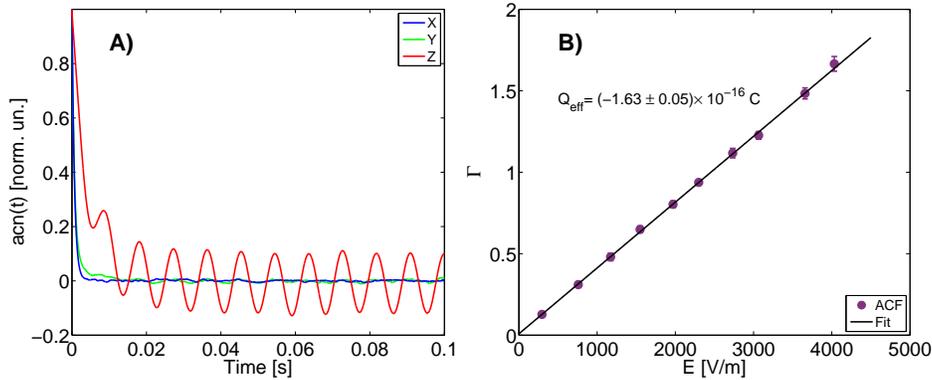}
\caption{A) Normalized autocorrelation function of the particle motion along the three axis. The oscillating electric field is applied 
along the z-axis. B) $\Gamma$ as function of the applied field amplitude $E_0$. The 
errorbars are the standard deviation over 10 repeated measurement for each field strength and are hardly visible, indicating 
the high accuracy of the technique and the high stability of the setup}
\label{fig.gamfield}
\end{center}
\end{figure}

In this experiment, we chose a modulation frequency $ f_d=96.7\,Hz$ which optimize the response of the trapped particle to the 
external field (for more details see  \cite{PesceE13}). A typical autocorrelation function for the three axis is shown in  Fig.\ref
{fig.gamfield}A, it is clearly visible the high signal to noise ratio achievable and that the oscillation of the particle occurs only 
along the z-axis where the uniform electric field was applied. From a fit of the cosinusoidal part of the autocorrelation function we 
measured the force ratio $\Gamma$. The behavior of the force ratio with the amplitude of the applied electric field  is shown in 
Fig.\ref{fig.gamfield}B.
The data were fitted with a straight line whose slope is the ratio $\Gamma/E$, then using Eq. (\ref{eq.charge}) we can 
evaluate the value of the effective charge of the polystyrene bead. The effective charge found was $Q_{eff}=
(-1.63\pm0.05)\times 10^{-16}\, C$, that compared to the value declared by the manufacturer ($Q=-1.79\times 10^{-13}\, 
C$) denotes that the condensation of free ions on the bead surface reduces the total charge of about one thousand times. 
The linear behavior of $\Gamma$ as function of the applied strength confirms that the charge is constant, i.e. the 
applied electric field doesn't affect it even if $\Gamma$ becomes greater than one where the electrophoretic force begins 
to dominate over the thermal force.

\section{Results and Discussion} 

\subsection{Measurement of the hydrodynamic coefficient of bacterial spores}
\label{sec:exp1}

\begin{figure}[t]
\begin{center}
\includegraphics[width=0.9\linewidth]{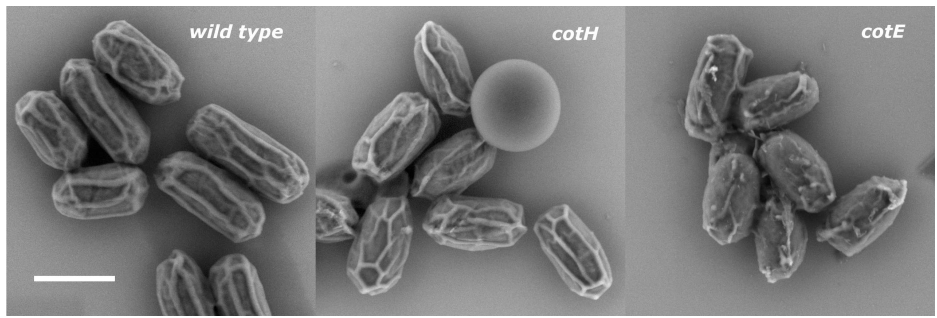}
\caption{SEM images of the spores used in this experiment. In the central picture a charged PS bead is also visible. Scale bar is 1 
$\mu m$}
\label{fig.sporesem}
\end{center}
\end{figure}

Figure \ref{fig.sporesem} shows a SEM image of the three spore types, used in this experiment, dried on a silicon substrate. 
The spores are similar in shape and their size is quite heterogeneous. They have a quasi-elliptical shapes but the surface is not 
smooth. The three spore types used in this study are all isogenic to the {\it wild type} strain PY79 but lack either the {\it cotE} 
(BZ213) or the {\it cotH} (ER220) gene because of gene replacements  \cite{ZhengGD88,NaclerioJB96}. As a consequence, 
spores of strain BZ213 totally lack the outer coat layer  \cite{ZhengGD88} while those of strain ER220 have a strongly defective 
coat, with several deviations with respect to the {\it wild type} at both the outer and inner coat levels  \cite{ZilhaoJB99}. Based on these 
differences it is not surprising that length and width, measured from SEM images for a large number of spores (N$>$100) of each 
strain, varies between the three spore types with the PY79 ({\it wild type}) being the largest and the BZ213 ({\it cotE}) the smallest 
(Table \ref{tab.spore}).
The values obtained for {\it wild type} spores are in good agreement with those reported by Huang et al. 
\cite{HuangV10}. 

\begin{table}[htdp]
\begin{center}
\begin{tabular}{lccc}
\hline
&{\it wild type}&{\it cotH}&{\it cotE} \\ \hline
\textbf{$a~ [\mu m]$}			&  1.42$\pm$0.09  &  1.26$\pm$0.08  &  1.16$\pm$0.09     \\ 
\textbf{$b~ [\mu m]$}			&  0.69$\pm$0.03  &  0.61$\pm$0.03  &  0.60$\pm$0.04      \\ 
\textbf{$\phi=a/b$}		&  2.1$\pm$0.2       &  2.1$\pm$0.2      &  1.9$\pm$0.2               \\ 
\end{tabular}
\end{center}
\caption{Spores diameters and aspect ratios. Errors are the standard deviations over 100 different particles.\label{tab.spore}}
\end{table}

Spores are stably trapped by the optical tweezers and, from an optical bright light image, it seems that their major axis is aligned along the beam propagation axis (z axis) as it can be seen in Fig.\ref{fig.setup}C where, for comparison, the image of a trapped microsphere is shown too. All the measurements (charge and hydrodynamic coefficient) were performed along this axis. Due to the quite large distribution of the spore sizes it was necessary to measure the hydrodynamic coefficient for each spore before evaluating its charge.

\begin{figure}[t]
\begin{center}
\includegraphics[width=0.8\linewidth]{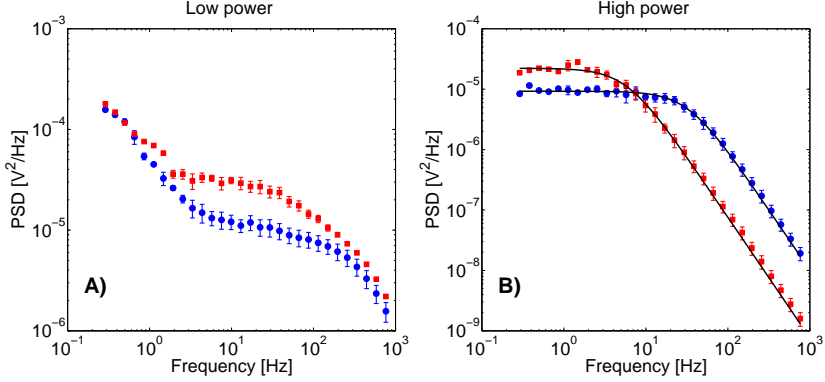}
\caption{The PSDs of a trapped wild type spore for two different powers. (A) At low power the motion of the bead is both rotational and translational and the PSD has a complex shape. (B) At high power the rotational motion disappears and the PSDs have a lorentzian shape as expected for a purely translational motion. Black solid lines are fitting to . For clarity only the motion along the x (blue) and z (red) axes are reported.}
\label{fig.psdmotion}
\end{center}
\end{figure}

To study the spore motion inside the optical potential well we recorded the motion of a trapped wild type spore at different trapping powers. We observed that at low power the spore displaces and rotates, as can be easily seen from the shape of the PSDs in 
Fig.\ref{fig.psdmotion}A. In fact, the PSDs along the three axes show different relaxation frequencies due to translational, angular and cross-correlated motions \cite{MaragoNL08}.  Increasing the trapping laser power, we observed a reduction of the angular motion until a certain value of the power ($\approx$6mW at the sample). Beyond that value, the motion was purely translational and the shape of the PSDs was a single lorentzian profile (see Fig.\ref{fig.psdmotion}B).

In this condition we have also checked for a possible presence of residual rotational effects due to the non-conservative forces always present in an optical trap. However, following the analysis described in a previous paper \cite{PesceEPL09}, we did not observe any evidence of these effects.

To validate the measurement procedure, we first trapped a polystyrene bead in the middle of the sample cell at a height of
$70\div80\, \mu m$  to avoid any influence from the walls, then we applied along the z-axis a sinusoidal oscillation to the 
PZT stage with amplitude $A=150$ nm and frequency $ f_d=8$ Hz  that is low enough to be sure that the fluid moves 
together with the stage. We recorded several trajectories of the particle position to measure $\beta$ and $\gamma$. 
This allowed us to check the goodness of the procedure since these two parameters can be also evaluated giving the 
viscosity of the water as known and extracting them from the result of a fitting of the PSD. 

\begin{figure}[t]
\begin{center}
\includegraphics[width=0.6\linewidth]{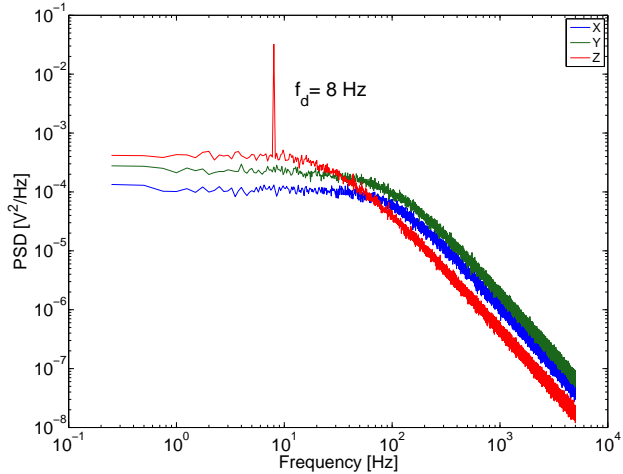}
\caption{PSDs along the three axis, the curves are slightly shifted to show the absence of peaks in the transverse 
plane.}
\label{fig.psdacf}
\end{center}
\end{figure}

Typical PSD functions obtained from this procedure are reported in Fig.\ref{fig.psdacf}. 
It is evident that the oscillation occurs only along the z-axis and is completely absent in the other directions.
The very flat plateau at low frequencies shows that the system is very stable. 
The sharp peak, in the PSD relative to z axis, at $f=8$ Hz is due to the fluid oscillation and is composed by 
one datum. This is achieved choosing a measurement time $T_m$ equal to an integer multiple $n$ of the 
oscillation period that is $T_m=n/ f_d$. In our case $T_m=20$ s so far we recorded exactly 160 oscillation 
periods. The signal power is simply the  area of the peak at $ f_d$ in the PSD and since it consists of a single 
point it follows that $W_{exp}=P_h \cdot \delta f$ where $P_h$ is the height of the peak datum with respect 
to the thermal noise plateau and $\delta f=1/T_m$ is the frequency resolution of the PSD.

From the data reported above, we obtained a value of $\gamma=(9.02\pm0.07)\times 10^{-9}
~N s/m$ (error is the standard deviation over 20 independent recordings). The well known Stokes' law 
\cite{Batchelor67} for  a sphere of diameter $d=1\,\mu m$ dispersed in a fluid of viscosity $\eta$ gives a 
value of the hydrodynamic coefficient equal to $\gamma=3\pi\eta d= 8.9\times 10^{-9}$, where the 
viscosity value of water at $T=22.7~^{\circ}C$ was $\eta=9.39\times10^{-4}~Pa\,s$  \cite{ChoJPCB99}. 
The measured value is in excellent agreement with the expected value.

As further test to evaluate the precision of the procedure, we measured the hydrodynamic coefficient of 100 different beads.
The ensemble average resulted: $\bar{\gamma}=9.1\pm0.12\,\times 10^{-9}~N\cdot s/m$. The value is, again, in good 
agreement with the expected value. Nevertheless it is worth to note that the uncertainty of 15\% is larger than that obtained 
for  the single bead measurement, because of the poly-dispersion of the beads solution (8\% as reported from the 
manufacturer), but also to other sources of noise (slow vibrations, electronics, temperature drifts).
These results confirm the accuracy and the goodness of the technique.

Once the technique has been validated with spherical particles, we were allowed to measure the hydrodynamic coefficient of 
spores.  To do that we recorded the trajectory of 100 spores and calculated the hydrodynamic coefficients. The average results 
for the different kinds of spores here used, are reported in Table\ref{tab.spore1} (together with the value obtained for the charged PS 
beads for comparison).  The expected values for the spores $\gamma_{th}$ were computed considering each spore as an ellipsoidal 
particle with cylindrical symmetry. As discussed above, in this case, the hydrodynamic coefficient is essentially the same of a sphere 
with a radius equal to the minor radius of the ellipsoid multiplied by a correction factor (see Eq. (\ref{eq.gfactor})). For aspect 
ratios around 2  like those used in this experiment (see Table \ref{tab.spore}), the correction is of about 20\%.

The comparison between expected and experimental values of $\gamma$, for the three spore types, shows very relevant discrepancies. The first result is that $\gamma_{exp}$ coefficients are all close but quite larger than the expected values. This could be ascribed to different reasons: (i) the spore surface is not smooth but rugged and very irregular as clearly visible in Fig.\ref{fig.sporesem}, (ii) the sporesÕ shape can be only approximately considered ellipsoidal, and (iii) even if the motion is purely translational, the spore could not be perfectly aligned and the velocity field of the fluid, during the oscillatory motion, can increase the alignment further leading to an higher {\it effective} drag coefficient.
More interesting results come from the comparison among investigated sporesÕ types.
A first result is that the values for {\it wild type} and {\it cotH} spores (see Table \ref{tab.spore1}) are similar although their minor diameters are different. In fact, despite the differences observed by TEM between PY79 and ER220 spores  \cite{ZilhaoJB99}, the final shape and surface appearance of spores of these two isogenic strains seem to be very similar. This conclusion is also supported by a functional analysis of spores of the ER220 strain, phenotypically indistinguishable from the isogenic wild type for the resistance to lysozyme treatment and the induction of germination \cite{NaclerioJB96}.

Another result is that the coefficient for {\it cotE} spores is significantly larger than the other two types of spores. This is an unexpected and very interesting result because since {\it cotE} spores have the short axis diameter smaller than {\it wild type} spores, should have a drag coefficient smaller. On the other hand, although {\it cotE} size is very close to that of {\it cotH} spores, again, their $\gamma_{exp}$ coefficient is rather larger. This suggests that other surface properties could influence the hydrodynamic behavior of spores. In fact, looking at the SEM image of Fig.\ref{fig.sporesem}, it is clear that the surface of {\it cotE} spores is drastically different with respect to the others. In addition, the almost total removal of the outer coat in {\it cotE} spores changes the spore size, but could also change the hydrophobicity of the surface as well. A decrease of the hydrophobicity could imply an increase of the friction between the spore surface and the surrounding fluid.

\begin{table}[htdp]
\begin{center}
\begin{tabular}{lcccc}
\hline
&{\it beads} &{\it wild type}	& {\it cotH} &{\it cotE}\\ \hline
\textbf{$\gamma_{exp}~[10^{-9}Ns/m]$} 	&	9.1$\pm$0.5 & 9.2$\pm$0.9     &  9.4$\pm$1.6     &  12$\pm$3         \\
\textbf{$\gamma_{th}~[10^{-9}Ns/m]$}	& 8.9$\pm$0.6 & 7.4$\pm$0.3     &  6.5$\pm$0.4   &  6.4$\pm$0.4             \\ 
\hline
\end{tabular}
\end{center}
\caption{Hydrodynamic coefficients of the beads and spores used in this experiment. Values in $10^{-9}~N s/m$.  The errors on the expected values come from the errors on the measurements of the size for the spores and from that reported by the manufacturer for the beads. For the expected values they come from the errors in the size measurement (see Table \ref{tab.spore}) in the case of spores, while for beads from the value reported by the manufacturer.\label{tab.spore1}}
\end{table}

\subsection{Charge measurement of bacterial spores}

Each type of spores, used in this work,  presents a certain degree of heterogeneity in shape and charge. So far, to measure their 
surface charge it was necessary to measure simultaneously, the force ratio and the hydrodynamic coefficient of a large number of 
different spores. Therefore we analysed 100 different spores for each type and polystyrene beads for comparison. 
For each of them we measured the hydrodynamic coefficient and the effective charge as described in the previous sections.
Every particle was trapped in the middle of the sample cell height, then, while the stage was oscillating along the z axis at $ f_d=8\,Hz
$ and $A=150\,nm$  we measured the hydrodynamic coefficient $\gamma$. Once this measurement was completed, with the stage at 
rest, we applied a periodic electric field at $ f_p=96.7\,Hz$ and  strength $E_0=1930\,V/m$  to measure the force ratio $\Gamma$ and then we calculated the value of the charge. The averaged results for all the spore type are reported in Table \ref{tab.charge} along with the values obtained for the PS bead.

\begin{table}[htdp]
\begin{center}
\begin{tabular}{lcccc}
\hline
            	&{\it beads}	 &{\it wild type}	&{\it cotH} & {\it cotE} \\ \hline
$Q_{eff}[10^{-16}\,C]$ 	&	-1.6$\pm$0.4	& -0.9$\pm$0.1     &  -1.2$\pm$0.1   &  -1.6$\pm$0.4 \\ 
$\zeta$   [mV]     &       -24.2$\pm$0.8 &    -30.7$\pm$0.9      & -28.4$\pm$0.7  & -34.7$\pm$0.8 \\  \hline
\end{tabular}
\end{center}
\caption{Values of the electric charge and zeta potential  for the beads and spores used in this experiment, we remember that all 
the charges measured are negative. Errors are the standard deviations over 100 different particles.}
\label{tab.charge}
\end{table}
The charge measured for the PS beads differs a little from that reported in the previous section. This is due to a small 
difference in the purity of water used in the two measurements. Moreover, these values are affected by an uncertainty 
larger than that obtained for a single particle measurement. This is due to the heterogeneity of the sample. Nevertheless it is 
interesting to note that for PS beads the uncertainty on the average measurement 
of $\gamma$ is in reasonable agreement with the poly-dispersion in size reported by the manufacturer. This suggests that PS 
beads have a very narrow distribution in size and shape but a quite wide distribution in charge.

Looking at the charge values of spores it is clear that the two mutants, {\it cotH} and {\it cotE} have a charge larger 
than the {\it wild type}. This suggest that removing the protein layers the surface charge increases. 
In literature it is reported that the surface of {\it B.subtilis} spores consists essentially of hexosamine-peptide organised in a manner 
that both saccharide and peptide components are exposed  \cite{DouglasTFS59,KazakovJPCB08}. The hexosamine-peptide can also 
contain both free carboxyl and free amino groups. Therefore the total charge is determined by the balance of these oppositely charged 
groups. The removal of outer coat proteins could modify this balance and then the total charge of the spore, but also the hydrophobicity 
properties of the surface.

As further comparison we measured the zeta potential of spores and bead with a commercial instruments (Zetasizer Nano 
ZS, Malvern) obtaining the values reported in Table \ref{tab.charge} at pH=6.5.
For a spherical particle of radius $a$ the potential is related to the total effective charge by the following relation:
\begin{equation}
\zeta=\frac{Q_{eff}}{4\pi\epsilon_0\epsilon_r a(1+ka)}
\label{eq.zetapot}
\end{equation}
where $k$ is the inverse of Debye length that for pure distilled water is equal to $k^{-1}=0.96\,\mu m$, this relation is valid 
in the limit of large particles (Helmoltz-Smoluchowski limit $ka\approx1$). Using Eq. (\ref{eq.zetapot}) for PS beads we found
$Q_{eff}=(-1.55\pm0.6)\times10^{-16}\,C$ which is, as expected, very close to the value obtained with our technique. 
It is difficult to compare the zeta potential values of spores because due to their irregular shape. 
An attempt can be done assuming an effective radius equal to: i) the average of the short and long radii of the spore, or ii) the 
long radius only.  The results, in the case of  {\it wild type} spore for example, are  $Q_{eff}=2.5\times10^{-16}\,C$ and 
$Q_{eff}=3.9\times10^{-16}\,C$, respectively. These values are of the same order of magnitude of that 
measured in this work, but not consistent with it and systematically larger. It worth to be noted that the value of zeta potential 
presented in this experiment are comparable to those reported in literature for the {\it wild type} spores \cite{HuangV10, 
AhimouJMM01}.

\section{Conclusions}

The effective charge of sulfate PS beads and of bacterial spores has been measured using optical tweezers. We have 
demonstrated that with this technique it is possible to discriminate between different mutants of bacterial spore by their effective 
charge. It is also possible to measure the hydrodynamic coefficient for particle with irregular shape, like spores, with high precision.
The technique allows the study of the heterogeneities in large samples in spite of the averaged results provided by most  
commercial instruments like zetasizers operating with conventional electrokinetic techniques. Moreover the presented method 
is rapid and reliable and can be easily extended to microfluidic devices.

\section*{Acknowledgment}        
The authors are very grateful to  Dr. Giovanni Romeo for measurements of the zeta potential.


\section*{References}

\begin{thebibliography}{34}
\expandafter\ifx\csname natexlab\endcsname\relax\def\natexlab#1{#1}\fi
\providecommand{\url}[1]{\texttt{#1}}
\providecommand{\href}[2]{#2}
\providecommand{\path}[1]{#1}
\providecommand{\DOIprefix}{doi:}
\providecommand{\ArXivprefix}{arXiv:}
\providecommand{\URLprefix}{URL: }
\providecommand{\Pubmedprefix}{pmid:}
\providecommand{\doi}[1]{\href{http://dx.doi.org/#1}{\path{#1}}}
\providecommand{\Pubmed}[1]{\href{pmid:#1}{\path{#1}}}
\providecommand{\bibinfo}[2]{#2}
\ifx\xfnm\relax \def\xfnm[#1]{\unskip,\space#1}\fi
\bibitem[{Cutting et~al.(2009)Cutting, Hong, Baccigalupi, and
  Ricca}]{CuttingIRI09}
\bibinfo{author}{S.~F. Cutting}, \bibinfo{author}{H.~A. Hong},
  \bibinfo{author}{L.~Baccigalupi}, \bibinfo{author}{E.~Ricca},
\newblock \bibinfo{title}{Oral vaccine delivery by recombinant spore
  probiotics},
\newblock \bibinfo{journal}{Int. Rev. of Immunol.} \bibinfo{volume}{28}
  (\bibinfo{year}{2009}) \bibinfo{pages}{487--505}.
\bibitem[{McKenney et~al.(2013)McKenney, Driks, and
  Eichenberger}]{McKenneyNRM13}
\bibinfo{author}{P.~T. McKenney}, \bibinfo{author}{A.~Driks},
  \bibinfo{author}{P.~Eichenberger},
\newblock \bibinfo{title}{The bacillus subtilis endospore: assembly and
  functions of the multilayered coat},
\newblock \bibinfo{journal}{Nat. Rev. Microbiol.} \bibinfo{volume}{11}
  (\bibinfo{year}{2013}) \bibinfo{pages}{33--44}.
\bibitem[{Huang et~al.(2010)Huang, Hong, Tong, Hoang, Brisson, and
  Cutting}]{HuangV10}
\bibinfo{author}{J.-M. Huang}, \bibinfo{author}{H.~A. Hong},
  \bibinfo{author}{H.~V. Tong}, \bibinfo{author}{T.~H. Hoang},
  \bibinfo{author}{A.~Brisson}, \bibinfo{author}{S.~M. Cutting},
\newblock \bibinfo{title}{Mucosal delivery of antigens using adsorption to
  bacterial spores},
\newblock \bibinfo{journal}{Vaccine} \bibinfo{volume}{28}
  (\bibinfo{year}{2010}) \bibinfo{pages}{1021--1030}.
\bibitem[{Sirec et~al.(2012)Sirec, Strazzulli, Isticato, Felice, Moracci, and
  Ricca}]{SirecMCF12}
\bibinfo{author}{T.~Sirec}, \bibinfo{author}{A.~Strazzulli},
  \bibinfo{author}{R.~Isticato}, \bibinfo{author}{M.~D. Felice},
  \bibinfo{author}{M.~Moracci}, \bibinfo{author}{E.~Ricca},
\newblock \bibinfo{title}{Adsorption of beta-galactosidase of alicyclobacillus
  acidocaldarius on wild type and mutants spores of bacillus subtilis},
\newblock \bibinfo{journal}{Microb. Cell Fact.} \bibinfo{volume}{11}
  (\bibinfo{year}{2012}).
\bibitem[{Isticato et~al.(2013)Isticato, Sirec, Treppiccione, Murano, Felice,
  Rossi, and Ricca}]{IsticatoMCF13}
\bibinfo{author}{R.~Isticato}, \bibinfo{author}{T.~Sirec},
  \bibinfo{author}{L.~Treppiccione}, \bibinfo{author}{F.~Murano},
  \bibinfo{author}{M.~D. Felice}, \bibinfo{author}{M.~Rossi},
  \bibinfo{author}{E.~Ricca},
\newblock \bibinfo{title}{Non-recombinant display of the b subunit of the heat
  labile toxin of {\it Escherichia coli} on wild type and mutant spores of {\it
  Bacillus subtilis}},
\newblock \bibinfo{journal}{Microb. Cell Fact.} \bibinfo{volume}{12}
  (\bibinfo{year}{2013}) \bibinfo{pages}{98}.
\bibitem[{Israelachvili(2011)}]{Israelachvili11}
\bibinfo{author}{J.~Israelachvili}, \bibinfo{title}{Intermolecular and Surface
  Forces: Revised Third Edition}, \bibinfo{publisher}{Elsevier Science},
  \bibinfo{year}{2011}.
\bibitem[{Reiber et~al.(2007)Reiber, K{\"o}ller, Palberg, Carrique, Reina, and
  Piazza}]{ReiberJCIS07}
\bibinfo{author}{H.~Reiber}, \bibinfo{author}{T.~K{\"o}ller},
  \bibinfo{author}{T.~Palberg}, \bibinfo{author}{F.~Carrique},
  \bibinfo{author}{E.~R. Reina}, \bibinfo{author}{R.~Piazza},
\newblock \bibinfo{title}{Salt concentration and particle density dependence of
  electrophoretic mobilities of spherical colloids in aqueous suspension},
\newblock \bibinfo{journal}{J. Coll. Interf. Sci.} \bibinfo{volume}{309}
  (\bibinfo{year}{2007}) \bibinfo{pages}{315--322}.
\bibitem[{Ashkin et~al.(1986)Ashkin, Dziedzic, Bjorkholm, and Chu}]{AshkinOL86}
\bibinfo{author}{A.~Ashkin}, \bibinfo{author}{J.~Dziedzic},
  \bibinfo{author}{J.~E. Bjorkholm}, \bibinfo{author}{S.~Chu},
\newblock \bibinfo{title}{Observation of a single-beam gradient force optical
  trap for dielectric particles},
\newblock \bibinfo{journal}{Opt. Lett.} \bibinfo{volume}{11}
  (\bibinfo{year}{1986}) \bibinfo{pages}{288--290}.
\bibitem[{Ashkin(1997)}]{AshkinPNAS97}
\bibinfo{author}{A.~Ashkin},
\newblock \bibinfo{title}{Optical trapping and manipulation of neutral
  particles using lasers},
\newblock \bibinfo{journal}{Proc. Natl. Acad. Sci. USA} \bibinfo{volume}{94}
  (\bibinfo{year}{1997}) \bibinfo{pages}{4853--60}.
\bibitem[{Gittes and Schmidt(1998)}]{GittesOL98}
\bibinfo{author}{F.~Gittes}, \bibinfo{author}{C.~Schmidt},
\newblock \bibinfo{title}{Interference model for back-focal-plane displacement
  detection in optical tweezers},
\newblock \bibinfo{journal}{Opt. Lett.} \bibinfo{volume}{23}
  (\bibinfo{year}{1998}) \bibinfo{pages}{7}.
\bibitem[{Rohrbach(2005)}]{RohrbachPRL05}
\bibinfo{author}{A.~Rohrbach},
\newblock \bibinfo{title}{Stiffness of optical traps: quantitative agreement
  between experiment and electromagnetic theory},
\newblock \bibinfo{journal}{Phys. Rev. Lett.} \bibinfo{volume}{95}
  (\bibinfo{year}{2005}) \bibinfo{pages}{168102}.
\bibitem[{Imparato et~al.(2007)Imparato, Peliti, Pesce, Rusciano, and
  Sasso}]{ImparatoPRE07}
\bibinfo{author}{A.~Imparato}, \bibinfo{author}{L.~Peliti},
  \bibinfo{author}{G.~Pesce}, \bibinfo{author}{G.~Rusciano},
  \bibinfo{author}{A.~Sasso},
\newblock \bibinfo{title}{Work and heat probability distribution of an
  optically driven brownian particle: Theory and experiments},
\newblock \bibinfo{journal}{Phys. Rev. E} \bibinfo{volume}{76}
  (\bibinfo{year}{2007}) \bibinfo{pages}{050101}.
\bibitem[{Strubbe et~al.(2008)Strubbe, Beunis, and Neyts}]{StrubbePRL07}
\bibinfo{author}{F.~Strubbe}, \bibinfo{author}{F.~Beunis},
  \bibinfo{author}{K.~Neyts},
\newblock \bibinfo{title}{Detection of elementary charges on colloidal
  particles},
\newblock \bibinfo{journal}{Phys. Rev. Lett.} \bibinfo{volume}{100}
  (\bibinfo{year}{2008}) \bibinfo{pages}{218301}.
\bibitem[{Roberts et~al.(2007)Roberts, Wood, Frith, and Bartlett}]{SethJCP07}
\bibinfo{author}{G.~S. Roberts}, \bibinfo{author}{T.~A. Wood},
  \bibinfo{author}{W.~J. Frith}, \bibinfo{author}{P.~Bartlett},
\newblock \bibinfo{title}{Direct measurement of the effective charge in
  nonpolar suspensions by optical tracking of single particles},
\newblock \bibinfo{journal}{J. Chem. Phys.} \bibinfo{volume}{126}
  (\bibinfo{year}{2007}) \bibinfo{pages}{194503}.
\bibitem[{Semenov et~al.(2009)Semenov, Otto, Stober, Papadopoulos, Keyser, and
  Kremer}]{SemenovJCIS09}
\bibinfo{author}{I.~Semenov}, \bibinfo{author}{O.~Otto},
  \bibinfo{author}{G.~Stober}, \bibinfo{author}{P.~Papadopoulos},
  \bibinfo{author}{U.~F. Keyser}, \bibinfo{author}{F.~Kremer},
\newblock \bibinfo{title}{Single colloid electrophoresis},
\newblock \bibinfo{journal}{J. Colloid. Interf. Sci.} \bibinfo{volume}{337}
  (\bibinfo{year}{2009}) \bibinfo{pages}{260--4}.
\bibitem[{Han et~al.(2006)Han, Alsayed, Nobili, Zhang, Lubensky, and
  Yodh}]{SCIHan06}
\bibinfo{author}{Y.~Han}, \bibinfo{author}{A.~M. Alsayed},
  \bibinfo{author}{M.~Nobili}, \bibinfo{author}{J.~Zhang},
  \bibinfo{author}{T.~C. Lubensky}, \bibinfo{author}{A.~G. Yodh},
\newblock \bibinfo{title}{Brownian motion of an ellipsoid},
\newblock \bibinfo{journal}{Science} \bibinfo{volume}{314}
  (\bibinfo{year}{2006}) \bibinfo{pages}{626--30}.
\bibitem[{Han et~al.(2009)Han, Alsayed, Nobili, and Yodh}]{PREHan09}
\bibinfo{author}{Y.~Han}, \bibinfo{author}{A.~Alsayed},
  \bibinfo{author}{M.~Nobili}, \bibinfo{author}{A.~Yodh},
\newblock \bibinfo{title}{Quasi-two-dimensional diffusion of single ellipsoids:
  Aspect ratio and confinement effects},
\newblock \bibinfo{journal}{Phys. Rev. E}  (\bibinfo{year}{2009}).
\bibitem[{Buosciolo et~al.(2004)Buosciolo, Pesce, and Sasso}]{BuoscioloOC04}
\bibinfo{author}{A.~Buosciolo}, \bibinfo{author}{G.~Pesce},
  \bibinfo{author}{A.~Sasso},
\newblock \bibinfo{title}{New calibration method for position detector for
  simultaneous measurements of force constants and local viscosity in optical
  tweezers},
\newblock \bibinfo{journal}{Opt. Commun.} \bibinfo{volume}{230}
  (\bibinfo{year}{2004}) \bibinfo{pages}{357--368}.
\bibitem[{Berg-Sorensen and Flyvbjerg(2004)}]{SorensenRSI04}
\bibinfo{author}{K.~Berg-Sorensen}, \bibinfo{author}{H.~Flyvbjerg},
\newblock \bibinfo{title}{Power spectrum analysis for optical tweezers},
\newblock \bibinfo{journal}{Rev. Sci. Inst.} \bibinfo{volume}{75}
  (\bibinfo{year}{2004}) \bibinfo{pages}{594--612}.
\bibitem[{Tolic-Norrelykke et~al.(2006)Tolic-Norrelykke, Schaeffer, Howard,
  Pavone, Juelicher, and Flyvbjerg}]{TolicRSI06}
\bibinfo{author}{S.~F. Tolic-Norrelykke}, \bibinfo{author}{E.~Schaeffer},
  \bibinfo{author}{J.~Howard}, \bibinfo{author}{F.~S. Pavone},
  \bibinfo{author}{F.~Juelicher}, \bibinfo{author}{H.~Flyvbjerg},
\newblock \bibinfo{title}{Calibration of optical tweezers with positional
  detection in the back focal plane},
\newblock \bibinfo{journal}{Rev. Sci. Inst.} \bibinfo{volume}{77}
  (\bibinfo{year}{2006}).
\bibitem[{Pesce et~al.(2005)Pesce, Sasso, and Fusco}]{PesceRSI05}
\bibinfo{author}{G.~Pesce}, \bibinfo{author}{A.~Sasso},
  \bibinfo{author}{S.~Fusco},
\newblock \bibinfo{title}{Viscosity measurements on micron-size scale using
  optical tweezers},
\newblock \bibinfo{journal}{Rev. Sci. Inst.} \bibinfo{volume}{76}
  (\bibinfo{year}{2005}) \bibinfo{pages}{115105}.
\bibitem[{Zheng et~al.(1988)Zheng, Donovan, Fitz-James, and Losick}]{ZhengGD88}
\bibinfo{author}{L.~B. Zheng}, \bibinfo{author}{W.~P. Donovan},
  \bibinfo{author}{P.~C. Fitz-James}, \bibinfo{author}{R.~Losick},
\newblock \bibinfo{title}{Gene encoding a morphogenic protein required in the
  assembly of the outer coat of the bacillus subtilis endospore.},
\newblock \bibinfo{journal}{Gene Develop.} \bibinfo{volume}{2}
  (\bibinfo{year}{1988}) \bibinfo{pages}{1047--1054}.
\bibitem[{Naclerio et~al.(1996)Naclerio, Baccigalupi, Zilh{\~a}o, De~Felice,
  and Ricca}]{NaclerioJB96}
\bibinfo{author}{G.~Naclerio}, \bibinfo{author}{L.~Baccigalupi},
  \bibinfo{author}{R.~Zilh{\~a}o}, \bibinfo{author}{M.~De~Felice},
  \bibinfo{author}{E.~Ricca},
\newblock \bibinfo{title}{Bacillus subtilis spore coat assembly requires coth
  gene expression.},
\newblock \bibinfo{journal}{J. Bacteriol.} \bibinfo{volume}{178}
  (\bibinfo{year}{1996}) \bibinfo{pages}{4375--80}.
\bibitem[{Nicholson and Setlow(1990)}]{Nicholson90}
\bibinfo{author}{W.~Nicholson}, \bibinfo{author}{P.~Setlow},
\newblock \bibinfo{title}{Sporulation, germination and out-growth},
\newblock in: \bibinfo{editor}{C.~Hardwood}, \bibinfo{editor}{S.~Cutting}
  (Eds.), \bibinfo{booktitle}{Molecular biological methods for Bacillus},
  \bibinfo{publisher}{John Wiley and Sons}, \bibinfo{address}{Chichester,
  United Kingdom}, \bibinfo{year}{1990}, p. \bibinfo{pages}{391Ð450}.
\bibitem[{Isticato et~al.(2008)Isticato, Pelosi, Zilh{\~a}o, Baccigalupi,
  O.Henriques, Felice, and Ricca}]{IsticatoJB08}
\bibinfo{author}{R.~Isticato}, \bibinfo{author}{A.~Pelosi},
  \bibinfo{author}{R.~Zilh{\~a}o}, \bibinfo{author}{L.~Baccigalupi},
  \bibinfo{author}{A.~O.Henriques}, \bibinfo{author}{M.~D. Felice},
  \bibinfo{author}{E.~Ricca},
\newblock \bibinfo{title}{Cotc-cotu heterodimerization during assembly of the
  bacillus subtilis spore coat},
\newblock \bibinfo{journal}{J. Bacteriol.} \bibinfo{volume}{190}
  (\bibinfo{year}{2008}) \bibinfo{pages}{1267--1275}.
\bibitem[{Pesce et~al.(2013)Pesce, Lisbino, Rusciano, and Sasso}]{PesceE13}
\bibinfo{author}{G.~Pesce}, \bibinfo{author}{V.~Lisbino},
  \bibinfo{author}{G.~Rusciano}, \bibinfo{author}{A.~Sasso},
\newblock \bibinfo{title}{{Optical manipulation of charged microparticles in
  polar fluids.}},
\newblock \bibinfo{journal}{Electrophoresis} \bibinfo{volume}{34}
  (\bibinfo{year}{2013}) \bibinfo{pages}{3141--3149}.
\bibitem[{Zilh{\~a}o et~al.(1999)Zilh{\~a}o, Naclerio, Henriques, Baccigalupi,
  Moran, and Ricca}]{ZilhaoJB99}
\bibinfo{author}{R.~Zilh{\~a}o}, \bibinfo{author}{G.~Naclerio},
  \bibinfo{author}{A.~O. Henriques}, \bibinfo{author}{L.~Baccigalupi},
  \bibinfo{author}{C.~P. Moran}, \bibinfo{author}{E.~Ricca},
\newblock \bibinfo{title}{Assembly requirements and role of coth during spore
  coat formation in bacillus subtilis},
\newblock \bibinfo{journal}{J. Bacteriol.} \bibinfo{volume}{181}
  (\bibinfo{year}{1999}) \bibinfo{pages}{2631--2633}.
\bibitem[{Marag{\`o} et~al.(2008)Marag{\`o}, Jones, Bonaccorso, Scardaci,
  Gucciardi, Rozhin, and Ferrari}]{MaragoNL08}
\bibinfo{author}{O.~M. Marag{\`o}}, \bibinfo{author}{P.~H. Jones},
  \bibinfo{author}{F.~Bonaccorso}, \bibinfo{author}{V.~Scardaci},
  \bibinfo{author}{P.~G. Gucciardi}, \bibinfo{author}{A.~G. Rozhin},
  \bibinfo{author}{A.~C. Ferrari},
\newblock \bibinfo{title}{{Femtonewton force sensing with optically trapped
  nanotubes}},
\newblock \bibinfo{journal}{Nano Letters}  (\bibinfo{year}{2008}).
\bibitem[{Pesce et~al.(2009)Pesce, Volpe, De~Luca, Rusciano, and
  Volpe}]{PesceEPL09}
\bibinfo{author}{G.~Pesce}, \bibinfo{author}{G.~Volpe}, \bibinfo{author}{A.~C.
  De~Luca}, \bibinfo{author}{G.~Rusciano}, \bibinfo{author}{G.~Volpe},
\newblock \bibinfo{title}{{Quantitative assessment of non-conservative
  radiation forces in an optical trap}},
\newblock \bibinfo{journal}{EPL (Europhysics Letters)}  (\bibinfo{year}{2009}).
\bibitem[{Batchelor(1967)}]{Batchelor67}
\bibinfo{author}{G.~K. Batchelor}, \bibinfo{title}{An Introduction to Fluid
  Dynamics}, \bibinfo{publisher}{Cambridge Mathematical Library},
  \bibinfo{year}{1967}.
\bibitem[{Cho et~al.(1999)Cho, Urquidi, Singh, and Robinson}]{ChoJPCB99}
\bibinfo{author}{C.~Cho}, \bibinfo{author}{J.~Urquidi},
  \bibinfo{author}{S.~Singh}, \bibinfo{author}{G.~Robinson},
\newblock \bibinfo{title}{Thermal offset viscosities of liquid \mbox{H$_2$O},
  \mbox{D$_2$O}, and \mbox{T$_2$O}},
\newblock \bibinfo{journal}{J. Phys. Chem. B} \bibinfo{volume}{103}
  (\bibinfo{year}{1999}) \bibinfo{pages}{1991--1994}.
\bibitem[{Douglas(1959)}]{DouglasTFS59}
\bibinfo{author}{H.~W. Douglas},
\newblock \bibinfo{title}{Electrophoretic studies on bacteria. part
  5.-interpretation of the effects of ph and ionic strength on the surface
  charge borne by b. subtilis spores{,} with some observations on other
  organisms},
\newblock \bibinfo{journal}{Trans. Faraday Soc.} \bibinfo{volume}{55}
  (\bibinfo{year}{1959}) \bibinfo{pages}{850--856}.
\bibitem[{Kazakov et~al.(2008)Kazakov, Bonvouloir, and
  Gazaryan}]{KazakovJPCB08}
\bibinfo{author}{S.~Kazakov}, \bibinfo{author}{E.~Bonvouloir},
  \bibinfo{author}{I.~Gazaryan},
\newblock \bibinfo{title}{Physicochemical characterization of natural ionic
  microreservoirs: Bacillus subtilis dormant spores},
\newblock \bibinfo{journal}{J. Phys. Chem. B} \bibinfo{volume}{112}
  (\bibinfo{year}{2008}) \bibinfo{pages}{2233--2244}.
\bibitem[{Ahimou et~al.(2001)Ahimou, Paquot, Jacques, Thonart, and
  Rouxhet}]{AhimouJMM01}
\bibinfo{author}{F.~Ahimou}, \bibinfo{author}{M.~Paquot},
  \bibinfo{author}{P.~Jacques}, \bibinfo{author}{P.~Thonart},
  \bibinfo{author}{P.~G. Rouxhet},
\newblock \bibinfo{title}{Influence of electrical properties on the evaluation
  of the surface hydrophobicity of bacillus subtilis},
\newblock \bibinfo{journal}{J. Microbiol. Meth.} \bibinfo{volume}{45}
  (\bibinfo{year}{2001}) \bibinfo{pages}{119--26}.

\end{thebibliography}
\end{document}